\documentclass[aps,showpacs,superscriptaddress,12pt]{revtex4}
\usepackage{epsfig}
\usepackage{fleqn}
\usepackage{amssymb,amsmath}

\begin{document}        
\thispagestyle{empty}

\title{ Experimental Study of the Reaction $\mathbf{e^+e^-\to K_SK_L}$ 
in the Energy Range $\mathbf{\sqrt{s}=1.04 \div 1.38}$ GeV.}

\begin{abstract}
We present a measurement of the $e^+e^-\to K_SK_L$ cross section in
the energy range $\sqrt{s}=1.04 \div 1.38\,\mathrm{GeV}$. For the
energy $\sqrt{s}\geq 1.2\,\mathrm{GeV}$ the cross section exceeds
vector meson dominance model predictions with only $\rho(770)$,
$\omega(783)$, and $\phi(1020)$ mesons taken into account. 
%This excess is due to contributions from the higher vector meson
%states like $\rho(1600)$, $\omega(1400)$, and $\phi(1680)$.
Measured cross section
agrees well with previous measurements.
\end{abstract}

\date{June 1, 2006}

\author{M.~N.~Achasov}
\affiliation{Budker Institute of Nuclear Physics, Novosibirsk, 630090, Russia}
\affiliation{Novosibirsk State University, Novosibirsk, 630090, Russia}

\author{K.~I.~Beloborodov}
\email{K.I.Beloborodov@inp.nsk.su}
\affiliation{Budker Institute of Nuclear Physics, 
Novosibirsk, 630090, Russia}
\affiliation{Novosibirsk State University,
Novosibirsk, 630090, Russia}

\author{A.~V.~Berdyugin}
\affiliation{Budker Institute of Nuclear Physics, 
Novosibirsk, 630090, Russia}

\author{A.~V.~Bozhenok}
\affiliation{Budker Institute of Nuclear Physics, 
Novosibirsk, 630090, Russia}
\affiliation{Novosibirsk State University,
Novosibirsk, 630090, Russia}

\author{D.~A.~Bukin}
\affiliation{Budker Institute of Nuclear Physics, 
Novosibirsk, 630090, Russia}

\author{T.~V.~Dimova}
\affiliation{Budker Institute of Nuclear Physics, 
Novosibirsk, 630090, Russia}

\author{V.~P.~Druzhinin}
\affiliation{Budker Institute of Nuclear Physics, 
Novosibirsk, 630090, Russia}
\affiliation{Novosibirsk State University,
Novosibirsk, 630090, Russia}

\author{V.~B.~Golubev}
\affiliation{Budker Institute of Nuclear Physics, 
Novosibirsk, 630090, Russia}
\affiliation{Novosibirsk State University,
Novosibirsk, 630090, Russia}

\author{I.~A.~Koop}
\affiliation{Budker Institute of Nuclear Physics, 
Novosibirsk, 630090, Russia}
\affiliation{Novosibirsk State University,
Novosibirsk, 630090, Russia}

\author{A.~A.~Korol}
\affiliation{Budker Institute of Nuclear Physics, 
Novosibirsk, 630090, Russia}

\author{S.~V.~Koshuba}
\affiliation{Budker Institute of Nuclear Physics, 
Novosibirsk, 630090, Russia}

\author{A.~P.~Lysenko}
\affiliation{Budker Institute of Nuclear Physics, 
Novosibirsk, 630090, Russia}

\author{A.~V.~Otboev}
\affiliation{Budker Institute of Nuclear Physics, 
Novosibirsk, 630090, Russia}

\author{E.~V.~Pakhtusova}
\affiliation{Budker Institute of Nuclear Physics, 
Novosibirsk, 630090, Russia}

\author{S.~I.~Serednyakov}
\affiliation{Budker Institute of Nuclear Physics, 
Novosibirsk, 630090, Russia}
\affiliation{Novosibirsk State University,
Novosibirsk, 630090, Russia}

\author{Yu.~M.~Shatunov}
\affiliation{Budker Institute of Nuclear Physics, 
Novosibirsk, 630090, Russia}
\affiliation{Novosibirsk State University,
Novosibirsk, 630090, Russia}

\author{V.~A.~Sidorov}
\affiliation{Budker Institute of Nuclear Physics, 
Novosibirsk, 630090, Russia}

\author{Z.~K.~Silagadze}
\affiliation{Budker Institute of Nuclear Physics, 
Novosibirsk, 630090, Russia}
\affiliation{Novosibirsk State University,
Novosibirsk, 630090, Russia}

\author{A.~N.~Skrinsky}
\affiliation{Budker Institute of Nuclear Physics, 
Novosibirsk, 630090, Russia}

\author{A.~V.~Vasiliev}
\affiliation{Budker Institute of Nuclear Physics, 
Novosibirsk, 630090, Russia}
\affiliation{Novosibirsk State University,
Novosibirsk, 630090, Russia}

               %Vector-meson dominance

\maketitle

\section{Introduction}
The spectroscopy of light-quark vector mesons is still far from
completion. It is mostly because not all the processes are measured
with sufficient accuracy. One of the sources of new information on
vector meson spectroscopy is a process $e^+e^-\to K_SK_L$, into which
contribute isoscalar $\omega(783)$, $\phi(1020)$, and
isovector $\rho(770)$ resonances as well as their higher mass excitations.
Accurate measurement of the $e^+e^-\to K_SK_L$
cross section is also important because it is a part of the total
cross section of the electron-positron annihilation into hadrons,
which enters into calculations of the hadronic vacuum polarization 
contribution to the
anomalous magnetic moment of the muon and running electromagnetic coupling
constant at $Z$-boson mass $\alpha_{\mathrm{em}}(M_Z)$.

At present the $e^+e^-\to K_SK_L$ cross section is measured with an
accuracy of few per cent only in the narrow energy interval around
$\phi(1020)$ resonance \cite{SND-1,CMD2-1}. First measurements of
this reaction at higher energy were conducted with DM1 \cite{bkMane}
and OLYA \cite{olya} detectors. The DM1 measurement at DCI collider 
covers the energy range $\sqrt{s}=1.40\div2.18\,\mathrm{GeV}$
with a total integrated luminosity of $1.4\,\mathrm{pb}^{-1}$. The
OLYA measurements in the energy range
$\sqrt{s}=1.06\div1.40\,\mathrm{GeV}$ were carried out at VEPP-2M
collider with the total integrated luminosity of
$0.7\,\mathrm{pb}^{-1}$. In both experiments significant excess of the
$e^+e^-\to K_SK_L$ cross section over vector meson dominance (VMD)
model was observed. The latest measurement of the $e^+e^-\to K_SK_L$
cross section was done by CMD-2 detector \cite{CMD2-2} in the energy
range $\sqrt{s}$ from 1.05 up to $1.40\,\mathrm{GeV}$.

In this work the measurement of the $e^+e^-\to K_SK_L$ cross section
in the energy range from 1.04 up to $1.38\,\mathrm{GeV}$ is
reported. The experiment was carried out with SND detector at VEPP-2M
$e^+e^-$ collider. The analysis is based on the total integrated
luminosity of $9.1\,\mathrm{pb}^{-1}$.

\section{Detector and experiment}

The SND detector \cite{SND} operated at VEPP-2M $e^+e^-$ collider
complex from 1995 up to 2000. It was designed as a universal detector
for the studies of the decays of $\rho$, $\omega$, $\phi$ resonances as
well as the processes of  $e^+e^-$ annihilation into hadrons in the
energy range $\sqrt{s} = 0.40\div1.40\,\mathrm{GeV}$.

The main part of the SND detector is a three-layer scintillation
electromagnetic calorimeter consisted of 1632 counters with NaI(Tl)
crystals. The total thickness of the calorimeter for the particles
originating from the detector center is 13.4 radiation lengths. The
energy resolution of the calorimeter for photons is $\sigma _E/E =
4.2\% /\sqrt[4\,]{E(\mathrm{GeV})}$. The angular resolution is equal
to $\sigma_{\phi},\sigma_{\theta}\simeq1.5^\circ$. The solid angle
coverage is 90\% of $4\pi$. The charged particle tracks are measured
by ten-layer drift chamber system located inside the calorimeter.

In this analysis the data collected in the experimental runs of 1997
and 1999 are used. In 1997 two scans of the energy range from 0.96 up
to $1.38\,\mathrm{GeV}$ with an energy step of $10\,\mathrm{MeV}$ and the
total integrated luminosity of $6\,\mathrm{pb}^{-1}$ were performed. The
year 1999 scan was done in the energy range from 1.02 to
$1.34\,\mathrm{GeV}$ with a step of $10\,\mathrm{MeV}$ and a total
integrated luminosity of $3.1\,\mathrm{pb}^{-1}$. In this analysis the
points with $\sqrt{s}\geq1.04\,\mathrm{GeV}$ were considered.

\section{Data analysis}

The process 
\begin{equation}
e^+e^- \to K_S K_L
\label{kskl}
\end{equation}
was studied in the decay mode $K_S \to \pi^0\pi^0 \to 4\gamma$. The
$K_L$ mesons, due to their large decay length, which is much larger
than the the detector radius, and large nuclear interaction length in
the NaI(Tl) ($\sim 0.35\,\mathrm{m}$), do not produce any signal in
the detector in a significant part of events. Nuclear interaction of
the $K_L$ meson or its decay inside the detector produce energy
deposits in the calorimeter counters, which are interpreted by the
event reconstruction program as one or more photons. In both cases
appearance of charged particles in the tracking system is
improbable. Thus, for the analysis of the process (\ref{kskl}) only the
events with no charged particle tracks in the drift chamber are
considered.
The main background processes for the process under study are the
following:
\begin{equation} 
e^+e^- \to \omega \pi^0 \to \pi^0 \pi^0 \gamma, 
\label{omp0n} 
\end{equation}
\begin{equation} 
e^+e^- \to \phi (\gamma) \to \eta \gamma (\gamma) \to 3 \pi^0 \gamma (\gamma),
\label{etag} 
\end{equation}
\begin{equation}
e^+e^- \to \phi \gamma \to K_S K_L \gamma
\label{ksklg}
\end{equation}
The process (\ref{ksklg}) is a radiative ``return'' to the $\phi$
resonance due to emission of photon(s) by initial particles. These
photons are mostly emitted at small angle with respect to the beam
direction and thus are not detected. The process (\ref{etag}) is a
sum of the processes $e^+e^-\to\eta \gamma$ and $e^+e^-\to\eta
\gamma\gamma$ with additional photon emitted by initial particles.
Other background sources considered in the analysis are beam
background and cosmic particles.

Initial event selection is based on the following criteria:
\begin{itemize}
\item[-] $N_\gamma \geq 4$, where $N_\gamma$ is a number of
reconstructed photons;
\item[-] $N_c = 0$, where $N_c$ is a number of reconstructed charged
particles;
\item[-] events with a cosmic particle track reconstructed in the
calorimeter are rejected. The track in the calorimeter is a
group of calorimeter crystal hits positioned along a common straight
line.
\end{itemize}
The last requirement reduces the number of selected events by more
than a factor of two, almost completely rejecting the background from
cosmic particles.

\begin{figure}
\begin{minipage}{.48\textwidth}
\centerline{\includegraphics[width=\textwidth]{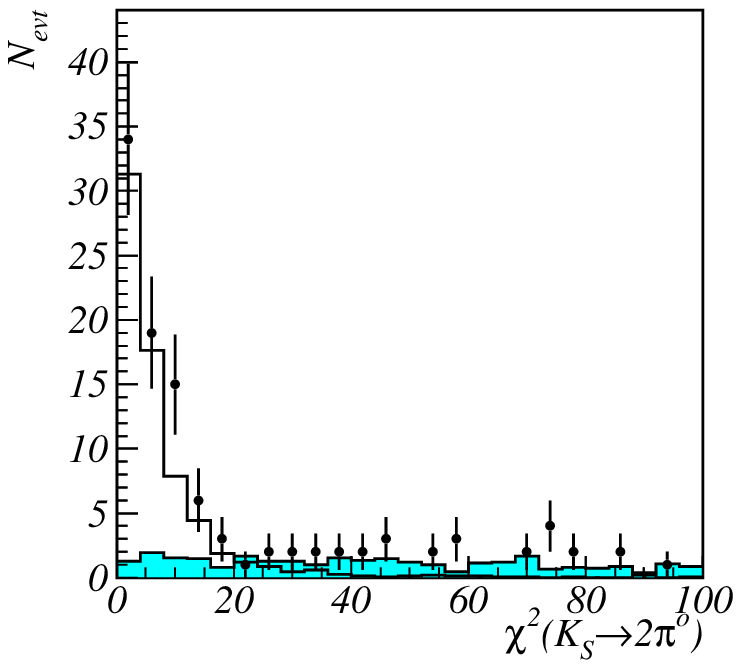}}
\caption{\label{xip0fonOmp} 
The $\chi^2_{K_S\to2\pi^0}$ distribution for the events with $\sqrt{s}
> 1.2\,\mathrm{GeV}$, $N_\gamma < 7$, and $E_{tot} \geq 0.5 \cdot
\sqrt{s}$.
Dots with error bars --- data, line --- simulation of the process 
(\ref{kskl}), shaded histogram --- simulation for the process
(\ref{omp0n}).}
\end{minipage}
\hfill
\begin{minipage}{.48\textwidth}
\centerline{\includegraphics[width=\textwidth]{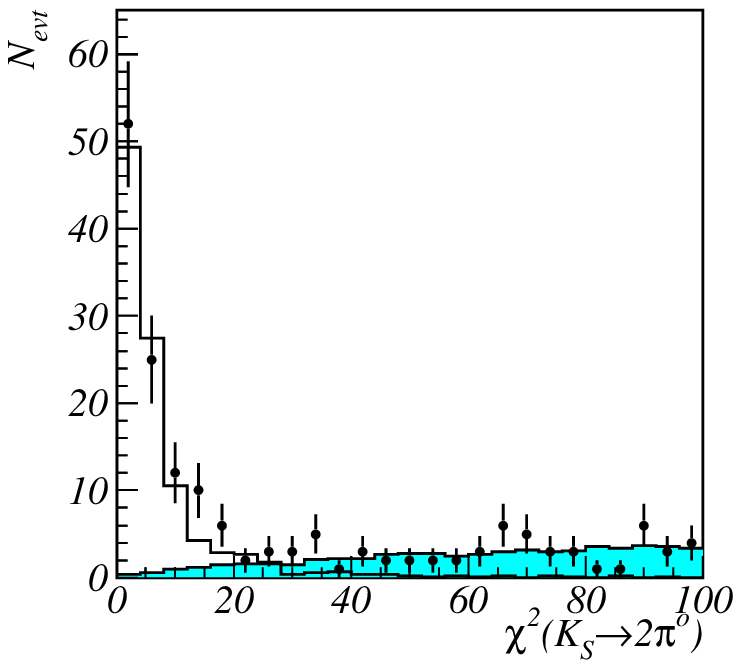}}
\caption{\label{xip0fonBKG}
The $\chi^2_{K_S\to2\pi^0}$ distribution for the events with $\sqrt{s}
> 1.12\,\mathrm{GeV}$ and $E_{tot} < 0.5 \cdot \sqrt{s}$.
Dots with error bars --- data, line --- simulation for the process
(\ref{kskl}), shaded histogram --- experimental distribution for the
beam background.}
\end{minipage}
\end{figure}

The events satisfying these criteria are kinematically fitted in the
hypothesis of $K_S \to \pi^0\pi^0 \to 4\gamma$ decay.
The kinematic fitting procedure
searches for the combination of two photon pairs from $\pi^0$ decays.
The invariant mass of found $\pi^0$'s is constrained to the mass of
$K_S$ meson. There is no constraints on $K_S$ energy. In a
multi-photon events for all four-photon
combinations corresponding $\chi^2$ are calculated, and the combination
with a minimum $\chi^2$ ($\chi^2_{K_S\to2\pi^0}$) is chosen. The
$\chi^2_{K_S\to2\pi^0}$ distributions for different selection criteria
are shown in Figs.~\ref{xip0fonOmp} and \ref{xip0fonBKG}. For further
analysis the events with $\chi^2_{K_S\to2\pi^0}<25$ are selected. The
following additional selection criteria are applied to these events:
\begin{enumerate}
\item $\zeta_i<0\, (i=1..4)$, where $\zeta_i$ is a ``quality'' parameter
of a reconstructed photon equal to $-\log L$, where
$L$ is a likelihood function value of a hypothesis
that observed transverse energy distribution in the cluster of hit calorimeter
crystals corresponds to one isolated photon~\cite{XINM}.
This parameter provides separation between events with isolated photon showers
and those with merged showers or clusters
from $K_L$-meson nuclear interactions or decays.
\item $36^\circ<\theta_i<144^\circ$, where $\theta_i$ is a polar
angle of the photon included in the reconstructed $K_S$ meson with
respect to the beam direction. This criterion rejects significant part
of the beam background.
\item $400 < M_{rec} < 550\,\mathrm{MeV}$, where $ M_{rec}$ is a
recoil mass of the reconstructed $K_S$ meson:
  \begin{equation}
     M_{rec}=\sqrt{ \sqrt{s} (\sqrt{s}-2E_{K_S})+M_{K_S}^2},
  \label{mrec}
  \end{equation}
$\sqrt{s}$ --- total energy in the center-of-mass frame, 
$E_{K_S}$ --- the energy of the reconstructed $K_S$ meson,
è $M_{K_S}$ is a $K_S$-meson mass. This requirement suppresses
background from the process (\ref{ksklg}). 
\item $\chi^2_{\pi^0\pi^0\gamma}> 60$, where
$\chi^2_{\pi^0\pi^0\gamma}$ is a $\chi^2$ of the kinematic fit in the
$e^+e^- \to \pi^0\pi^0\gamma$ hypothesis. This criterion is applied to
events with $N_{\gamma}\geq5$ to suppress the background from the
process (\ref{omp0n}).
\end{enumerate}

The total number of events satisfying all criteria described above in the
full energy range is equal to 1998, of which 585 events are in the region
$\sqrt s\geq 1.1\,\mathrm{GeV}$.
 
The number of background events of the process
$e^+e^-\to\omega\pi^0$ was estimated from Monte Carlo simulation. The
cross section of the process (\ref{omp0n})  was taken from the paper
\cite{OmP0}, in which was also shown that the simulation of the
multi-photon events in SND detector agrees with experiment within 5\%.
We checked the accuracy of the $e^+e^-\to\omega\pi^0$
background estimation using experimental data.
The error estimation is based on approximation of the
$\chi^2_{K_S\to2\pi^0}$ distribution (Fig.~\ref{xip0fonOmp}) for
events with $\sqrt{s} > 1.2\,\mathrm{GeV}$,
$N_\gamma < 7$ and $E_{tot} \geq 0.5 \cdot \sqrt{s}$, where $E_{tot}$
is a total energy deposition in the calorimeter. These requirements
effectively suppress contributions from all background processes
except (\ref{omp0n}). The distribution is approximated by a sum of the
process under study and (\ref{omp0n}), obtained by simulation and
shown in Fig.~\ref{xip0fonOmp}. This experimental estimation agrees
within statistical accuracy with that obtained by simulation.
The total number of events of this background process in the full
energy range with all selection criteria applied is
$N_{\omega\pi^0}=11.3\pm0.3\pm2.3$.

The number of background events of the process
$e^+e^-\to\eta\gamma(\gamma)$ is also estimated by simulation. The
events of this process satisfying all selection criteria are dominated
by radiative return to $\phi$ resonance:
$e^+e^-\to\phi\gamma,\,\phi\to\eta\gamma$, in which the photons emitted
by initial particles are not detected. The accuracy of the estimation
of this process contribution
is $\approx 3\%$. It is determined by the accuracy
of the $e^+e^-\to\eta\gamma$ cross section measurements near the
$\phi$ resonance ($\approx 2\%$), and the accuracy of the
$e^+e^-\to\eta\gamma\gamma$ cross section calculation
 ($1\%$). The total number of the background events of this
process is $21.2\pm0.2\pm1.1$.

For the estimation of the beam background the $\chi^2_{K_S\to2\pi^0}$
distribution shown in Fig.~\ref{xip0fonBKG} was used. The dots with
error bars show the experimental $\chi^2_{K_S\to2\pi^0}$ distribution
for the energy range $\sqrt{s}>1.12\,\mathrm{GeV}$ with additional
selection criterion $E_{tot}<0.5\sqrt{s}$, which rejects events of all
background processes except beam background. The shaded histogram
shows the experimental distribution for the beam background events,
for which the selection criteria inverse of 1 and 2 were used. The
total number of the beam background events is estimated as
$N_{b}=30\pm3\pm5$.

\begin{figure}
\centerline{\includegraphics[width=0.75\textwidth]{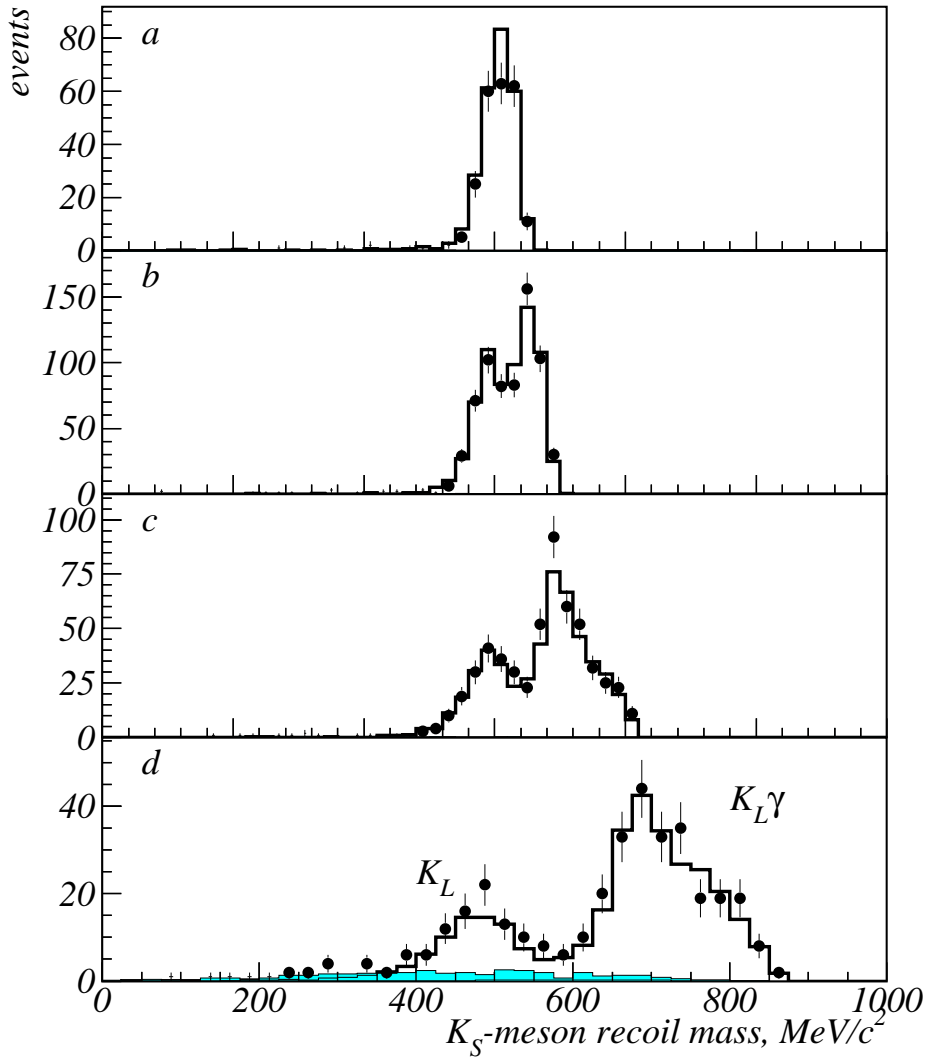}}
\caption{\label{mkrcn}
The recoil mass spectra against reconstructed $K_S$ meson in four
energy intervals: 
{\it a}) $\sqrt{s}=1.04-1.05\,\mathrm{GeV}$,
{\it b}) $\sqrt{s}=1.06-1.09\,\mathrm{GeV}$, 
{\it c}) $\sqrt{s}=1.10-1.20\,\mathrm{GeV}$,
{\it d}) $\sqrt{s}=1.20-1.38\,\mathrm{GeV}$.
The spectrum is for events, satisfying additional selection criteria:
$N_\gamma < 7$ è $E_{tot} \geq 0.5\sqrt{s}$. The shaded histogram
shows estimated $e^+e^-\to\omega\pi^0$ background.}
\end{figure}
The contribution of the $e^+e^-\to K_SK_L\gamma$ events strongly
depends on the beam energy. For the energy region close to $\phi$
resonance the photon energy $E_\gamma$ is small in comparison with the
total energy $\sqrt{s}$ making this process virtually
indistinguishable from $e^+e^-\to K_SK_L$. With the increase of collision
energy the photon energy grows and the kinematics of the
processes (\ref{kskl}) and (\ref{ksklg}) becomes more distinct.
The recoil mass spectra against reconstructed $K_S$ meson are shown in
Fig.~\ref{mkrcn} for four energy intervals. The peak with a mean value
close to the $K^0$-meson mass is due to to the reaction $e^+e^-\to
K_SK_L$, the rest corresponds to the process $e^+e^-\to
K_SK_L\gamma$. Good separation between these processes can be achieved
at energies above 1.2~GeV. At lower energy the processes
(\ref{kskl}) and (\ref{ksklg}) cannot be separated. To solve this
problem the approximation of the cross section was carried out with
the detection efficiency as a function of both $\sqrt{s}$ and the
energy of the photon emitted by initial particles.

The distribution of the number of selected events,
background from $e^+e^-\to\omega\pi^0$ and $e^+e^-\to\eta\gamma
(\gamma)$ as  function of energy
with the beam background subtracted is presented in
Table~\ref{table1}.

\section{Detection efficiency \label{detef}} 
\begin{figure}[tbp]
\begin{minipage}{0.45\textwidth}
\centerline{\includegraphics[width=0.9\textwidth]{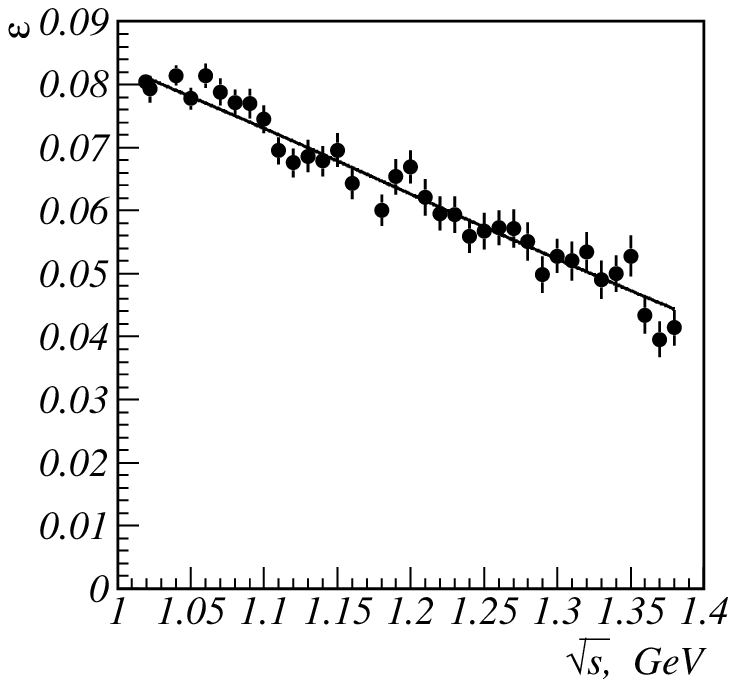}}
\caption{\label{eff1}
Detection efficiency for the process (\ref{kskl}) as a function of
energy for the events with the energy of the photon emitted by initial
particles $E_\gamma<10\,\mathrm{MeV}$.}
\end{minipage}
\hfill
\begin{minipage}{.45\textwidth}
\centerline{\includegraphics[width=0.9\textwidth]{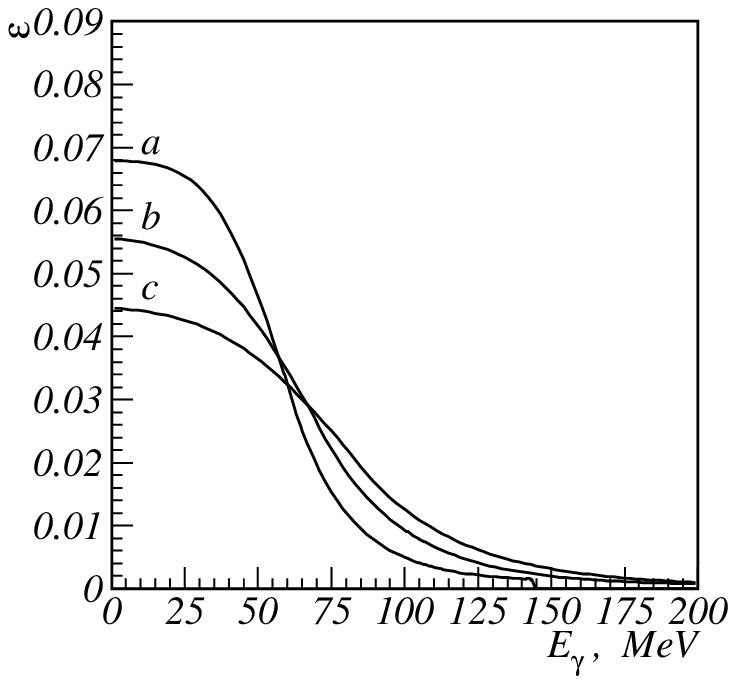}}
\caption{\label{eff2}
Detection efficiency for the process (\ref{kskl}) as a function of
the energy $E_\gamma$ of the photon emitted by initial particles for
three different energies:
{\it a} --- $\sqrt{s}=1.15\,\mathrm{GeV}$, 
{\it b} --- $\sqrt{s}=1.27\,\mathrm{GeV}$, 
{\it c} --- $\sqrt{s}=1.38\,\mathrm{GeV}$.
}
\end{minipage}
\end{figure}
The detection efficiency for the process under study was determined
using Monte Carlo simulation. The simulation takes into account photon
emission by initial particles \cite{RadCor, RadCor2}, permitting
us to take into account the dependence of the detection efficiency 
$\varepsilon(\sqrt{s},z)$ on energy $\sqrt{s}$ and
$z=\frac{E_\gamma}{\sqrt{s}}$, the fraction of energy
 carried away by the photon
emitted by initial particle. The energy dependence of the
detection efficiency for the process (\ref{kskl}) with
$E_\gamma<10\,\mathrm{MeV}$ is shown in Fig.~\ref{eff1}. The
decrease of the efficiency at large $\sqrt{s}$ is caused 
by degradation of resolution in kinematically fitted $K_S$-meson recoil mass for energies,
which are far from $K_S K_L$ production threshold.
%increase ofthe decay length of $K_S$ meson which is not taken into account in the kinematic fit. 
%This degrades the angular resolution for photons and
%finally leads to broadening of the $K_L$-meson mass peak in the
%$K_S$-meson recoil mass spectrum (Fig.~\ref{mkrcn}) at high energies.
Fig.~\ref{eff2} shows the detection efficiency dependence on
$E_\gamma$ for several energy points.

The correction for the detection efficiency, which takes into account
the difference in detector response between data and simulation was
estimated using events from the $\phi$-resonance region, where the
process (\ref{kskl}) can be separated with a negligible background
without constraints on $\chi^2_{K_S\to2\pi^0}$ and $\zeta_i$. The
following selection criteria were applied:
$N_{\gamma}=5$, $E_{\gamma_{5}} > 100\,\mathrm{MeV}$,
$\angle(\mathbf{n}_{K_S},\mathbf{n}_{\gamma_{5}}) > 130^{\circ}$, where
$E_{\gamma_{5}}$ is an energy of the photon not included in the
reconstructed $K_S$ meson, 
$\angle(\mathbf{n}_{K_S},\mathbf{n}_{\gamma_{5}})$ is an angle between
directions of the reconstructed $K_S$ meson and redundant photon. For
experimental and simulated events selected using these requirements
the fractions $r_{MC}$ and $r_{exp}$ of events satisfying standard
selection criteria were calculated. The correction for the
detection efficiency is determined as
$\kappa=r_{exp}/r_{MC}$. It is equal to $0.956\pm0.015$. The
quoted error is statistical.

\section{Determination of the Born cross section}
The visible cross section $\sigma_{vis}$ of the process under study,
which is directly obtained from experimental data, is related to Born
cross section $\sigma_0$ as:
\begin{equation}
\sigma_{vis}(\sqrt{s})=\int\limits^{1}_{0} dz\cdot\sigma_{0}
(\sqrt{s}(1-z))\cdot F(z,s)\cdot\varepsilon(\sqrt{s},z)
\label{bornsec}
\end{equation}
\noindent where $F(z,s)$ is a probability density function for the
initial particles to emit photon carrying the fraction of energy
$z\sqrt{s}$  \cite{RadCor}, $\varepsilon(\sqrt{s},z)$ is a detection
efficiency as a function of $\sqrt{s}$ and $z$. The experimental Born
cross section is determined using following procedure. The measured
visible cross section as a function of energy
$\sigma_{vis,i}={N_i}/{\mathit{IL}_i}$ (here $N_i$ is a number of
selected events after background subtraction for the $i$-th energy
point, $\mathit{IL}_i$ is an integrated luminosity for this point) is
approximated by a function calculated using Eq.~(\ref{bornsec}) with
some model for the Born cross section. As a result of the
approximation the parameters of this model are calculated together
with the function $R(s)=\sigma_{vis}(s)/\sigma_{0}(s)$. Experimental
values for the Born cross section are determined then according to the
following equation:
\begin{equation}
\sigma_{0,i}=\frac{\sigma_{vis,i}}{R(s_i)}.
\label{secborni}
\end{equation}
Model dependence of the result is estimated by variation of the Born
cross section models.

The Born cross section of the process $e^+e^-\to K_SK_L$ was
considered in a framework of the VDM:
\begin{equation}
\sigma_{0}(s)= \frac{12\pi}{s^{3/2}}
\left|\sum_{V=\rho,\omega,\phi,{\ldots}}
\frac{\sqrt{\Gamma_{V\to K_SK_L}(s)\Gamma_{V\to ee}m_V^3}e^{i\theta_V}}
{s-m_V^2+im_V \Gamma_V(s)}\right|^2  \label{vacfit}
\end{equation}

The ratios of partial widths and relative phases of $\rho$, $\omega$, and $\phi$
mesons were taken according to SU(3) model: 
\begin{center}
$\Gamma_{\rho\to K_SK_L}(s)=\Gamma_{\omega\to K_SK_L}(s)=2\Gamma_{\phi\to
K_SK_L}(s)$,
\end{center}
\begin{center}
$\theta_\rho=0^0$, $\theta_\omega=180^0$, $\theta_\phi=180^0$
\end{center}
The masses $m_V$ and full widths  $\Gamma_V$ for the excitations of
$\rho$, $\omega$ è $\phi$ were taken as in \cite{PDG}.
The approximation is done using program FIT \cite{pFIT}. The following
models for the Born cross section were considered:
\begin{enumerate}
\item{ The process is described by four vector mesons 
$\rho$, $\omega$, $\phi$ è $\rho(1450)$. The relative phase $\theta_{\rho(1450)}=0^{\circ}$,}
\item{ The process is described by four vector mesons
$\rho$, $\omega$, $\phi$ è $\phi(1680)$. The relative phase $\theta_{\phi(1680)}=0^{\circ}$,}
\item{ The process is described by four vector mesons
$\rho$, $\omega$, $\phi$ è $\rho(1700)$. The relative phase $\theta_{\rho(1700)}=0^{\circ}$.}
\end{enumerate}
For the models described above the following $\chi^2$ values were
obtained:
$\chi^2_1/ndf=19.1/21$, 
$\chi^2_2/ndf=18.9/21$ è $\chi^2_3/ndf=18.7/21$, respectively. All
three models provide good approximation of the experimental data and
can be used for the estimation of the Born cross section. Our final
result is based on approximation by a second model. The cross section
values are listed in Table~\ref{table1}.
\begin{table}
\caption{\label{table1}
Born cross section of the process $e^+e^-\to K_SK_L$, measured by SND detector.
$N_{exp}$ --- the number of selected events, $\sum N_{bkg}$ --- the
number of background events,
$\varepsilon_i$ --- detection efficiency, $1+\delta_i$ --- radiative
correction. The quoted errors are statistical and systematic, respectively}
\begin{center}
\begin{tabular}[t]{|c|r|r|r|r|r|r|}
\hline
$\sqrt{s}$, GeV&IL, nb$^{-1}$&$N_{exp}$&$\sum N_{bkg}$&
$\varepsilon_i$&$1+\delta_i$&$\sigma_0$, nb \\
\hline
$1.04$ & $69$ & $245$ & $3.0\pm1.0$ & $0.079$ & $1.61$ & $27.3\pm1.8\pm0.8$ \\
$1.05$ & $83$ & $183$ & $2.0\pm0.8$ & $0.078$ & $1.81$ & $15.5\pm1.2\pm0.5$ \\
$1.06$ & $274$ & $421$ & $2.5\pm1.1$ & $0.077$ & $1.92$ & $10.3\pm0.5\pm0.3$ \\
$1.07$ & $97$ & $96$ & $1.2\pm0.6$ & $0.076$ & $1.78$ & $7.2\pm0.8\pm0.2$ \\
$1.08$ & $572$ & $420$ & $6.2\pm1.3$ & $0.075$ & $1.49$ & $6.4\pm0.3\pm0.2$ \\
$1.09$ & $94$ & $48$ & $0.9\pm0.4$ & $0.074$ & $1.30$ & $5.2\pm0.8\pm0.2$ \\
$1.10$ & $436$ & $158$ & $4.8\pm0.9$ & $0.073$ & $1.20$ & $4.0\pm0.3\pm0.1$ \\
$1.11$ & $88$ & $21$ & $1.3\pm0.4$ & $0.072$ & $1.13$ & $2.75\pm0.65\pm0.08$ \\
$1.12-1.13$ & $420$ & $97$ & $5.6\pm1.0$ & $0.071$ & $1.09$ & $2.81\pm0.30\pm0.09$ \\
$1.14-1.15$ & $358$ & $61$ & $3.6\pm0.7$ & $0.069$ & $1.05$ & $2.22\pm0.30\pm0.07$ \\
$1.16$ & $316$ & $40$ & $2.3\pm0.5$ & $0.067$ & $1.02$ & $1.74\pm0.29\pm0.05$ \\
$1.18-1.19$ & $587$ & $44$ & $4.3\pm0.8$ & $0.065$ & $1.00$ & $1.04\pm0.18\pm0.03$ \\
$1.20-1.21$ & $569$ & $32$ & $3.7\pm0.7$ & $0.063$ & $0.99$ & $0.80\pm0.16\pm0.03$ \\
$1.22-1.23$ & $465$ & $25$ & $3.8\pm0.8$ & $0.060$ & $0.99$ & $0.77\pm0.18\pm0.02$ \\
$1.24-1.25$ & $562$ & $22$ & $2.2\pm0.5$ & $0.058$ & $0.98$ & $0.62\pm0.15\pm0.02$ \\
$1.26-1.27$ & $397$ & $16$ & $1.6\pm0.4$ & $0.056$ & $0.98$ & $0.66^{+0.24}_{-0.18}\pm0.02$ \\
$1.28-1.29$ & $492$ & $20$ & $2.4\pm0.6$ & $0.054$ & $0.97$ & $0.68^{+0.22}_{-0.17}\pm0.02$ \\
$1.30-1.31$ & $459$ & $11$ & $1.0\pm0.2$ & $0.052$ & $0.97$ & $0.43^{+0.19}_{-0.14}\pm0.01$ \\
$1.32-1.33$ & $516$ & $3$ & $2.1\pm0.5$ & $0.050$ & $0.97$ & $0.04^{+0.12}_{-0.07}\pm0.01$ \\
$1.34-1.35$ & $676$ & $13$ & $2.7\pm0.6$ & $0.048$ & $0.97$ & $0.33^{+0.15}_{-0.11}\pm0.01$ \\
$1.36$ & $606$ & $11$ & $3.1\pm0.7$ & $0.047$ & $0.96$ & $0.29^{+0.16}_{-0.12}\pm0.01$ \\
$1.37-1.38$ & $722$ & $11$ & $2.2\pm0.5$ & $0.045$ & $0.96$ & $0.28^{+0.14}_{-0.11}\pm0.01$ \\
\hline					     
\end{tabular}
\end{center}
\end{table}
Also shown in the table are the values of the detection efficiency
and radiative corrections calculated as: 
\begin{equation}
1+\delta(s)=\frac{\int\limits^{1}_{0} dz\cdot\sigma_{0}(\sqrt{s}(1-z))\cdot F(z,s)}{\sigma_{0}(s)}
\label{radcorr}
\end{equation}
\begin{equation}
\varepsilon(s)=\frac{\sigma_{vis}(s)}{\sigma_{0}(s)\cdot (1+\delta(s))}
\label{eff}
\end{equation}
The function $R(s)$ introduced above is defined as
$R(s)=\varepsilon(s)(1+\delta(s))$.

\section{Systematic errors}
The full systematic error on Born cross section of the process
$e^+e^-\to K_SK_L$ includes several contributions: the integrated
luminosity uncertainty, the uncertainty in the detection efficiency,
the errors on the beam subtraction and radiative corrections
estimations. 

{\bf The error on integrated luminosity.} The integrated luminosity at
SND detector is determined using QED processes $e^+e^-\to e^+e^-$ and
$e^+e^-\to\gamma\gamma$, for which the cross sections are known with
a precision of better than 1\%. As an estimate of the systematic error
we take the difference between the luminosities obtained using these
processes, which is about 2\% almost independently of the beam energy.

{\bf The uncertainty of the detection efficiency}
The analysis of the systematic uncertainty of the detection efficiency
was done using events from the $\phi$-resonance energy region. The
cross section of the process (\ref{kskl}) was measured using events
with four or more reconstructed photons. The events of this class
contain four photons from the $K_S\to2\pi^0$ decay and additional
clusters from decays or nuclear interactions of $K_L$ mesons or
clusters from beam background. As it was described above the selection
of the events of the process (\ref{kskl}) is based on reconstruction
of $K_S$ meson. No constrains on additional photons were applied. Such
approach minimizes the systematic error from the inaccuracy of the
simulation of the nuclear interaction of $K_L$ meson in the
detector. Nevertheless, since it is not possible to distinguish the
clusters from the $K_S$-meson decays from those from those produced by
$K_L$ meson or beam background, these additional clusters give rise to
a combinatorial background where lost or misreconstructed photons from
$K_S$ decay are replaced by clusters produced by $K_L$ or beam
background. The more such clusters, the larger number of
misreconstructed $K_S$ mesons. This effect leads to an uncertainty of
the detection efficiency due to inaccuracy of simulation of the
$K_L$-meson interactions in the detector material. This includes not
only the inaccuracy of the total cross section of the nuclear
interaction of $K_L$ meson, but also the inaccuracy of the number of
clusters and energy depositions in the calorimeter counters.

The total systematic error on detection efficiency is a sum
of the uncertainty for the ``pure'' $K_S$ meson and uncertainty of the
combinatorial background increased by inaccuracy of the simulation of
the nuclear interaction of $K_L$ meson in the detector.

The uncertainty of the simulation of the ``pure'' $K_S$ meson was
studied using events, in which the $K_L$ meson is reconstructed as a
single photon. The selection criteria for these events are described
in section \ref{detef}. It was found that corresponding correction to
the efficiency estimated by simulation is equal to $0.956\pm
0.015$. This correction accounts for the differences in distributions
in $\chi^2$ (3\%) and photon quality parameter (2\%).

The systematic error on combinatorial background and nuclear
interaction of $K_L$ mesons was estimated as a difference between the
detection efficiency values obtained by two ways. 
The first, standard,  way is to estimate detection efficiency
$\varepsilon_{MC}$ as the ratio of the numbers of selected signal events 
to the total number of simulated events.
 The second way is to
first estimate the detection efficiencies for the subsets of simulated
events with fixed numbers of reconstructed photons and then to
average obtained values according to relative weights observed
in data. The ratio of the efficiencies obtained in this way
is $\varepsilon_{MC}^\ast/\varepsilon_{MC}=0.991\pm0.007$.
Corresponding systematic uncertainty is equal to 1\%.

Another source of systematic uncertainty is the difference in energy and angular resolutions 
for photons between data and simulation. This difference affects the resolution in recoil mass
of the reconstructed $K_S$ meson shown in Fig.~\ref{mkrcn} and in the slope of the dependence
of the detection efficiency of the process (\ref{kskl}) on $E_{\gamma}$, shown in Fig.~\ref{eff2}.
To evaluate this contribution into the systematic error we varied the slope of the efficiency
dependence on $E_\gamma$ within limits corresponding to observed 2\% difference in  
$K_S$ recoil mass resolution in data and simulation on $\phi$ resonance.

The total systematic error on detection efficiency changes from 2.1\% to 2.5\% with the energy
in the range 1.04 -- 1.4~GeV.

\begin{table}
\caption{\label{table3}
Systematic uncertainties of the measured Born cross section of the
process $e^+e^-\to K_SK_L$.}
\begin{center}
\begin{tabular}[t]{|l||c|}
\hline
 Source & 1.02 --- 1.40 GeV \\
\hline
 Integrated luminosity & 2\%\\
 Detection efficiency & 2.1---2.5\%\\
 Background subtraction &0.1---2.9\%\\
 Model dependence&0.5 --- 3.0\%\\
\hline
 Total&2.9 --- 5.3\%\\
\hline
\end{tabular}
\end{center}
\end{table}

{\bf Background subtraction.}
As it was mentioned above, the contributions of background processes
\ref{omp0n} and \ref{etag} well agree with with estimations based on
simulation. Therefore the extraction of the Born cross section from
the experimental data the estimated by simulation number of background
events of these processes was subtracted. The beam background was
estimated from the $\chi^2_{K_S\to2\pi^0}$ distribution of the
experimental events. The systematic error on the background
subtraction varies from 0.1\% to 2.9\% for the energy range from 1.04
up to 1.40~GeV. The statistical error on the background subtraction is
included into the quoted statistical error of the measured cross
section.

{\bf Accuracy of the radiative corrections}
This systematic error includes the theoretical uncertainty of the
radiative correction calculation, which does not exceed
0.1\%~\cite{RadCor}, and model dependence, related to the choice of
the model describing the energy dependence of the cross section being
measured. As an error estimate the difference between results obtained
with the different choices of the approximation function described in
section V. The systematic error from radiative correction uncertainty
varies between 0.5\% and 3.0\% in the energy range from 1.04 to 1.40~GeV.

The total systematic errors are listed in Table~\ref{table3}.

\section{Discussion}

\begin{figure}[tbp]
\centerline{\includegraphics[width=0.9\textwidth]{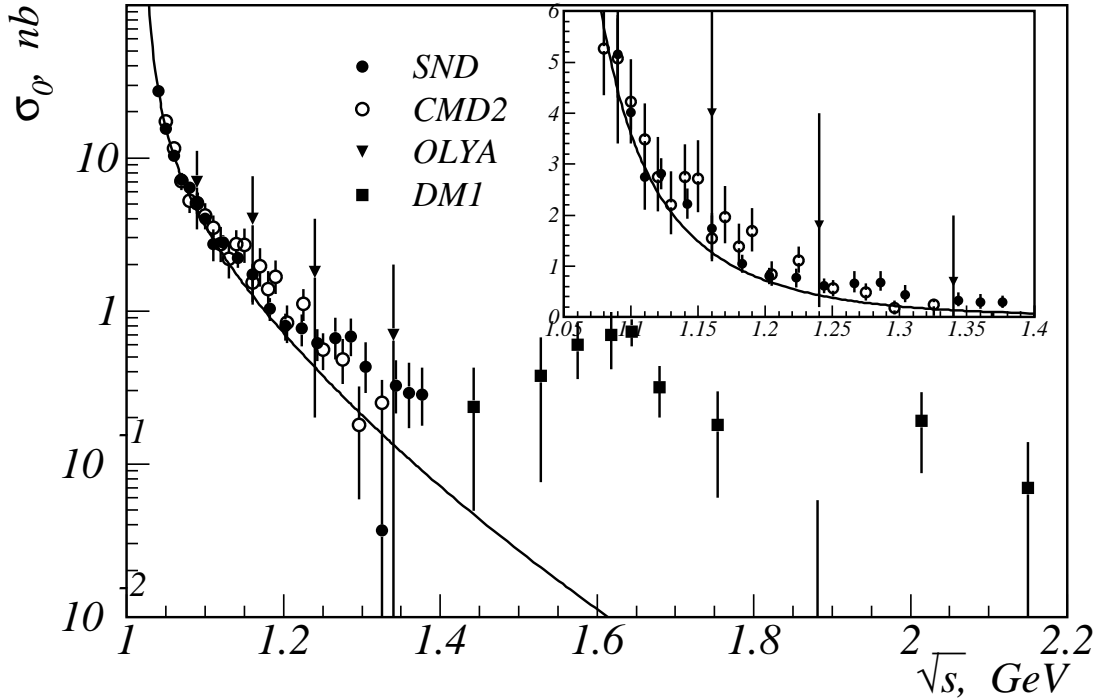}}
\caption{\label{crsecb}
         Born cross section of the process $e^+e^-\to K_SK_L$.
	 Dots with error bars  represent results obtained by 
	 SND(this work), CMD-2\cite{CMD2-2}, OLYA\cite{olya} and
         DM1\cite{bkMane} detectors. 
	 Solid line represents the cross section calculated using 
	 VDM with $\rho(770)$, $\omega(783)$, $\phi(1020)$ taken into
         account.
}
\end{figure}

In Fig.~\ref{crsecb} the cross sections measured in this work and
previous measurements by OLYA, CMD-2, and DM1 detectors are shown. The
results are in good agreement with ones obtained by CMD-2 and OLYA in
the same energy region.

The measured cross section of the process $e^+e^-\to K_SK_L$
significantly exceeds one predicted by VMD model with only $\rho(770)$,
$\omega(783)$, and $\phi(1020)$ resonances considered within SU(3) model.
The curve corresponding to this prediction is shown in Fig~\ref{crsecb}.
The measured cross section significantly exceeds this VMD estimation
starting from the energy about 1.2~GeV. This excess can be described by
contributions from higher mass states $\rho^{\prime}$, $\omega^\prime$, and
$\phi^\prime$, but for the determination of the parameters of these states
the data in wider energy range and for other decay modes are needed.
 
This work was partially supported by the RFFI grants 05-02-16250-a,
04-02-16181-a, 04-02-16184-a and the Sci.School-905.2006.2.

\clearpage

\end{document}